\documentclass[10pt]{article}
\usepackage{arxiv}
\usepackage[utf8]{inputenc} 
\usepackage[T1]{fontenc}    
\usepackage{hyperref}       
\usepackage{url}            
\usepackage{booktabs}       
\usepackage{amsfonts}       
\usepackage{nicefrac}       
\usepackage{microtype}      
\usepackage{lipsum}		
\usepackage{graphicx}
\usepackage{soul}
\usepackage{subcaption}
\usepackage{ccicons}
\usepackage{CJKutf8}
\usepackage{amsmath}

\hypersetup{%
  pdftitle={Automatic Photo to Ideophone Manga Matching},
  pdfauthor={David A. Shamma, Tony Dunnigan, Lyndon Kennedy},
  pdfkeywords={ideophone, mimetic, onomatopoeic, manga, comic, photo,
  GloVe, annotation},
  bookmarksopen=false,
  pdfstartview={FitH},
  hidelinks=true,
  colorlinks=true,
  citecolor=black,
  filecolor=black,
  linkcolor=black,
  urlcolor=black,
  breaklinks=true,
}

\title{Automatic Photo to Ideophone Manga Matching}
\newcommand{\authorname}[2]{\href{#1}{\hspace{-2.5mm}\includegraphics[width=2.5mm]{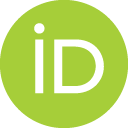}\hspace{1mm}#2}}

\author{%
  \authorname{https://orcid.org/0000-0003-2399-9374} 
  {David A.~Shamma} 
  \\ 
  FXPAL\\
  Palo Alto, CA, USA \\
  \texttt{aymans@acm.org} \\
  \And%
  \authorname{https://orcid.org/0000-0002-6233-794X} 
  {Tony Dunnigan} \\
  FXPAL\\
  Palo Alto, CA, USA \\
  \texttt{tony@dunnigan.net} \\
  \And%
  \authorname{https://orcid.org/0000-0002-1756-836X} 
  {Lyndon Kennedy} \\
  FXPAL\\
  Palo Alto, CA, USA \\
  \texttt{lyndonk@acm.org} \\
}

\renewcommand{\headeright}{Technical Report}
\renewcommand{\undertitle}{Technical Report}

\begin{document}

\maketitle

\begin{abstract}
Photo applications offer tools for annotation via text and stickers.  Ideophones, mimetic and onomatopoeic words, which are common in graphic novels, have yet to be explored for photo annotation use. We present a method for automatic ideophone recommendation and positioning of the text on photos. These annotations are accomplished by obtaining a list of ideophones with English definitions and applying a suite of visual object detectors to the image. Next, a semantic embedding maps the visual objects to the possible relevant ideophones. Our system stands in contrast to traditional computer vision-based annotation systems, which stop at recommending object and scene-level annotation, by providing annotations that are communicative, fun, and engaging. We test these annotations in Japanese and find they carry a strong preference and increase enjoyment and sharing likelihood when compared to unannotated and object-based annotated photos.
\end{abstract}


\section{Introduction}
Photo annotations in camera and messaging applications are on the
rise. This includes a wide variety from geo-based recommendations
(like the game score for a photo taken in a sports arena) to rich
augmented reality on faces in the picture. Beyond this, annotations
have little relation to the current photo's visual content.
Currently, some systems can recommend an annotation as the object name
or matching sticker of that exact object for manual placement on the
photo, this still does not capture the expression and visual nature of
ideophones and onomatopoeia that is seen in graphic novels, manga, and
comic strips.

In this work, we present a system for automatically predicting and
placing onomatopoeic annotations on photos based on the content of the
photos themselves. This is accomplished by applying a suite of visual
object detectors to the photo and then mapping to a dictionary of
onomatopoeic terms through an intermediate semantic embedding, wherein
the onomatopoeic terms (in Japanese) are represented by their
explanations (in English) and compared against the object recognition
vocabulary (also in English).

\begin{figure}
  \centering
  \begin{subfigure}[t]{0.30\columnwidth}
    \includegraphics[width=\columnwidth]{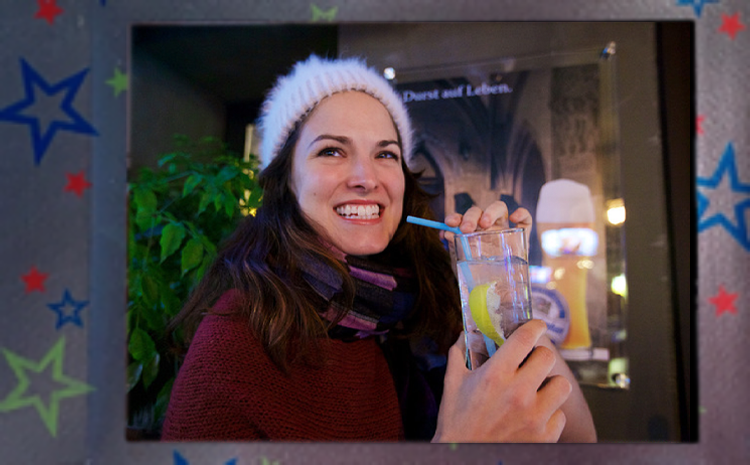}
    \caption{Source photo.}\label{fig:tease:src}
  \end{subfigure}
  \hspace{0.5pc}
  \begin{subfigure}[t]{0.30\columnwidth}
    \includegraphics[width=\columnwidth]{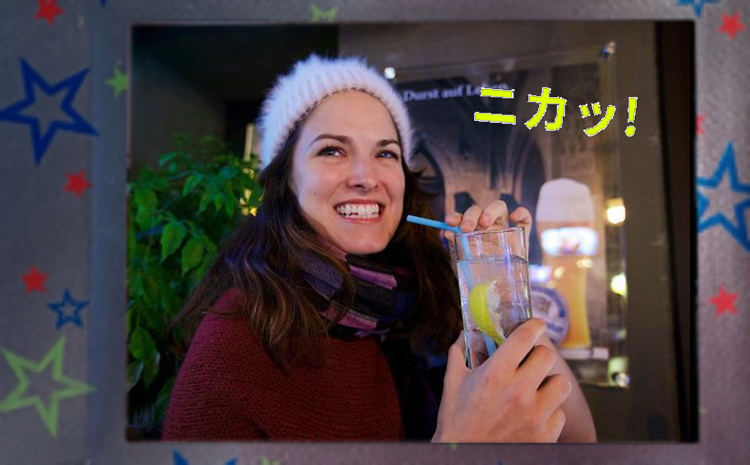}
    \caption{Annotated photo.}\label{fig:tease:annote}
  \end{subfigure}   
  \caption{\label{fig:teaser} A source photo and an photo
    automatically annotated with the Japanese Ideophone for smile.}
\end{figure}

We compare this system against a baseline of simply suggesting
annotations that are directly drawn from object and scene
classification engines. While these baseline annotations are very much
aligned with the semantic content of the images, we find that users
overwhelmingly prefer mimetic and onomatopoeic annotations for the
purposes of sharing photos for communicative purposes. This suggests
that the task of annotating photos with semantic tags is not aligned
with many users' practice of using photos for communication purposes.

\section{Related Work}
For decades, a large amount of computer vision research has focused on
identifying objects~\cite{790410,gevers1999color} (such as
cars, bikes, and people) and other contextual
factors~\cite{zhou2014learning,xiao2010sun} (like grass, sky, and
pavement) within images. Recent advances in deep
learning~\cite{krizhevsky2012imagenet} and increases in the
availability of training data~\cite{Thomee:2016:YND:2886013.2812802}
have made these applications ultimately very feasible, to the extent
that they are readily available in many commercial applications (such
as Google Photos or Flickr). Simultaneously, users have gained access
to increasingly powerful smartphones, with the capability of capturing
and immediately sharing high quality photographs. While these
smartphone applications enable the overlay of any arbitrary label,
users rarely want to just state the semantic content of the image (for
example: ``train'') and much prefer to use the communicative or
evocative meaning (``chug, chug, chug'').

This assumption is informed by user-centered
work~\cite{Ames:2007:WWT:1240624.1240772} about the behavior of users
tagging their own photos and using those tags for communication
purposes. Largely, photographs contain semantic content, but the
purpose of tagging is often for communicating contextual factors, such
as location, time, and experiences. This, again, suggests that while
semantic image tagging might be important for retrieval applications,
it might not be appropriate for sharing and communication.

There has been some work on Manga-style generation such as using the
device accelerometer or other methods to detect movement or silence
and render the mimetic term~\cite{Umeda_2012,Matsumura_2013} or to
stylize (comic-ize) the image~\cite{Fujieda_2017} without the term.
There's a host of work around Computational Manga and
Anime~\cite{Wong_2013} which also deals with style and composition but
not ideophones or mimetic text.  Perhaps the most related work are the
sticker recommendations from the SnapChat application.  These are
geographic and time based recommendations for stickers which are to be
manually placed on the photo, resized, and saved. The geo-lookup finds
the temperature outside and the movement speed of the camera in
addition to several events like a holiday or a local sports score.

\section{Photo to Ideophone Matching}
For the photo to ideophone matching, we propose a fully automated
approach where a user takes a photograph with a camera or app (or
selects a photo from a collection) and the annotation is recommended
and automatically composited and rendered on the photo or print. This
requires three steps: an initialization, a matching, then an
execution.  This process can be optimized to live embedded on a device
but could be cloud-connected as well.

\subsection{Initialization}\label{sec:setup}
\begin{CJK}{UTF8}{min}
  A dictionary of ideophones is needed.  We selected theJADEDnetwork's
  open-source community driven dictionary with a large Japanese
  ideophone to English definition repository with 1,329 contextual
  manga related terms~\cite{web:JADED}.  A typical example would be:
  \textsf{トケー}$\Rightarrow$\textit{the sound of a mechanical
    clock's internal mechanism}. In effect, any language, or icon
  even, for the term can be used, however the definition needs to be
  in English.  Here, the English term would be \textsf{tic-toc}.  The
  dictionary from theJADEDnetwork provides Japanese as a matching pair
  of Hirigana and Katakana, romanji, several English equivalents, and
  an explanation (the definition). For example: ジーッ, じーっ; ji-;
  (1) *whine* (2) *stare*; (1) Like when microphone is too close to
  the speakers, see also *Ui-n*; (2) As in staring at someone, or
  looking at something for an extended period of time. Comes from the
  ``ji-'' in ``jiro jiro miru'' (じろじろ見る).
  
  Next for each definition, we construct a score based on a term
  vector created using a GloVe~\cite{pennington2014glove}.
  \begin{equation}
    \frac{\sum_{i=1}^n\mathrm{GloVe}(t_i)}{n} \label{glovevec}
  \end{equation}
  Where $\mathrm{GloVe(t)}$ is the GloVe score for the given term $t$.
  We selected the pretrained Wikipedia 2014 + Gigaword 5 vector (6B
  tokens, 400K vocab, uncased, 50d) as the model to use.
  For every term in our dictionary, this becomes 1,329 50 dimension
  score vectors.
\end{CJK}

\subsection{Matching}
With the dictionary model complete, we need to construct a similar
photo vector. Given any photo, we run a set of visual classifiers.  We
use SqueezeNet as the object classifier as well as a food classifier
with 2000 dishes based on MobileNet, as well as, a face and smile
detector.~\cite{i2016squeezenet,howard2017mobilenets,web:aiy} Each
classifier returns a set of objects $o$ and confidence scores for
those objects $c_o$. The smile classifier returns a floating-point
score from $0.0\ldots1.0$ where $0.0$ is a frown and $1.0$ is a
smile. For the sake of consistency, we report this as a smile for
anything $>=0.5$ normalized from $0.0\ldots1.0$. Similarly, a frown is
scored the same way for scores $<0.5$ normalized inversely (so a smile
score of $0.1$ would map to a frown at $0.9$).  For each classifier,
we compute a GloVe vector similar to Eq~\eqref{glovevec} however, we
weight each object by its detection confidence as:
\begin{equation}
  \frac{\sum_{j=1}^n c_{o_j} \mathrm{GloVe}(o_j)}{n} \label{gloveviz}
\end{equation}
This results in three 50 dimensional vectors. For each classifier
vector, we find the top 5 minimal Cosine distance between itself and
the ideophone dictionary GloVe vectors from \S\ref{sec:setup}. Now,
the closest vector can be picked and the term can be recommended.  We
returned the top 5 in the previous step to allow for some jitter to
prevent a single term from being printed repeatably.

\subsection{Compositing}\label{subsec:exec}
Photo and term in hand, compositing the mimetic term is the next
step. This requires knowing where to place it so not to disturb the
image or occlude the subject.  There are several methods to find
salient (and inversely non-salient) regions in an image particularly
from the multimedia
literature~\cite{Mao:2012:MDS:2393347.2396302,Ye:2014:MCS:2578726.2578738,Tang:2017:SOD:3123266.3123318}.
For this system, we opted to implement a simplified method for
non-salient region detection.  First, using OpenCV's findCountours
function, we find largest contour in the image.  We then split the
image a quadrant plane at the midpoint. Then, we find the quadrant
with the smallest intersection to the largest contour. With a target
quadrant, the term is composited and rendered in the photo's corner at
a random size and angle,both at an empirically threshold to be mostly
level and not too large, onto the photo. The final output is produced
(see Figure~\ref{fig:tease:annote} for an example).

\subsection{Fabrication}
The initial camera system was put into a custom fabricated camera for
field testing. A Raspberry Pi Zero with a camera was used with the
Google AIY~\cite{web:aiy} accelerator kit (for running computer vision
models).  The Ideophone GloVe vectors were sparse compressed using
python3 and the vector matching is computed on the CPU.\@ The term
matching takes 30 seconds on average while the computer vision modules
take about 90 seconds (as each model has to be loaded and unloaded as
the device carries limited memory).  Finally, the photo is printed via
WiFi to an Instax SP2 mobile photo printer.  The whole system is self
contained and needs no network uplink or interface aside from the
camera shutter actuator.
\begin{figure}
  \centering
  \includegraphics[height=0.25\columnwidth]{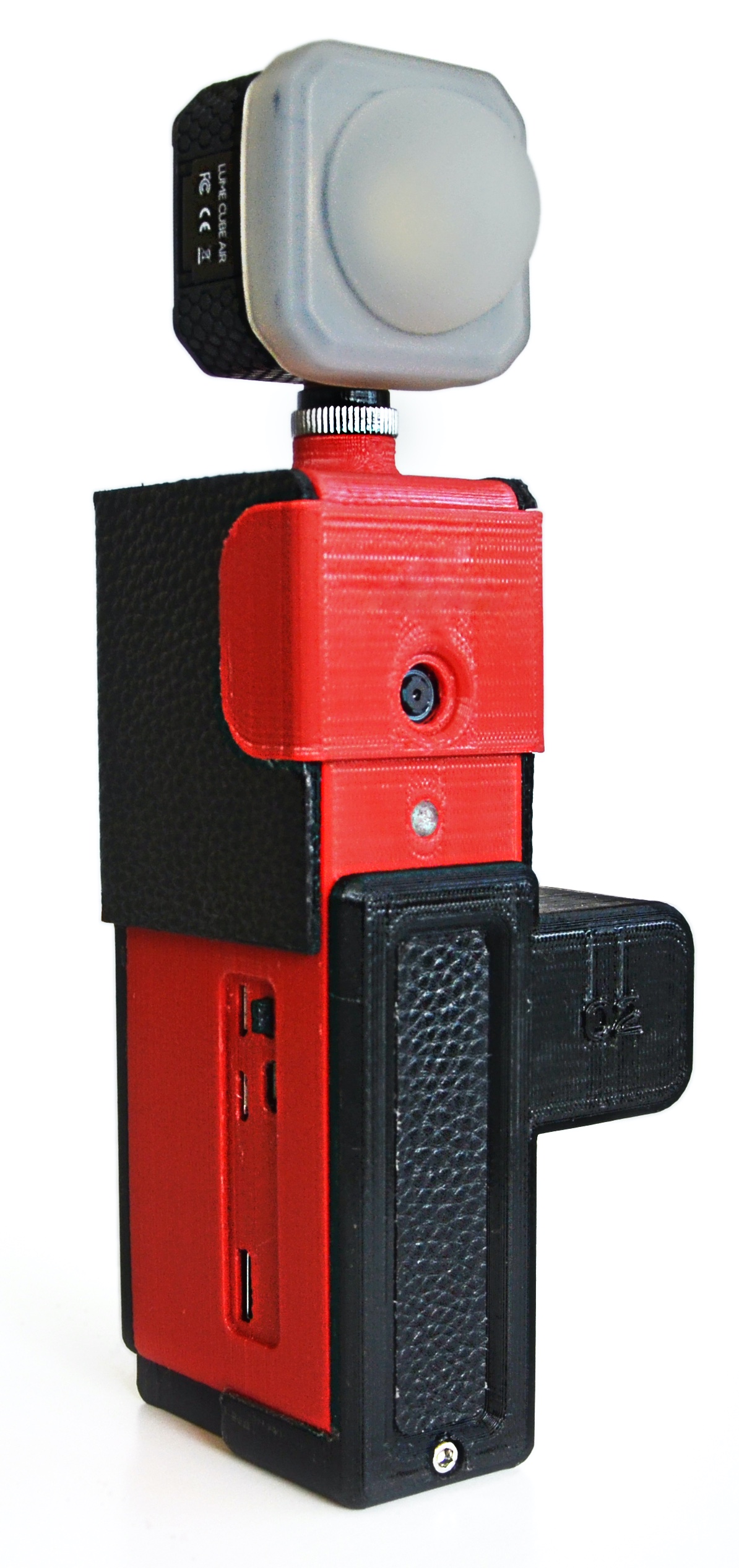}
  \caption{\label{fig:camera} The 3D printed Ideophone camera made
    with Raspberry Pi and Google AIY accelerator. }
\end{figure}

\section{Evaluation}
We begin with a preliminary evaluation of the ideophone matching
method.  A larger experiment would be needed to test people on their
own photos with either our fabricated camera (see
Figure~\ref{fig:camera}) or a camera app on a mobile phone to measure
how people feel about mimetic annotations on their personal photos and
photo messages.  For the preliminary evaluation, we needed to collect
photos. Here we used top terms from the YFCC100M
dataset~\cite{Thomee:2016:YND:2886013.2812802} to select a sample of
photos across a variety of topics and quality levels. Photos were
chosen from the dataset or from the personal collection of the authors
that matched the search for common terms like cat, smile, lightning.
Several photos were selected from the YFCC100M dataset to ensure
aesthetic coverage. For example, a scan of some money and a receipt.
This was done to represent a personal camera roll as photo content
types tend to vary on personal
devices.~\cite{Boulanger:2016:DPP:2901790.2901810}

Next, we designed a short survey to gauge how people felt about a set
of photos, how they compared these photos to their annotated versions
(object term versus mimetic term), and finally a side by side
comparison of two annotated photos. The survey was written in English
and Japanese; we used English for testing the survey with 5 people and
ran the survey with 10 participants (60\% female) in the Tokyo Metro
area.  Fluent and proficient Japanese language readers were recruited
using a third party service and participants were paid 15 USD.\@ The
ages ranged from 28 to 50 with a median of 33. Half of the
participants were Japanese and all could read hiragana, katakana, and
kanji.  The survey was completed in an average of 23 minutes.  For
photography experience, all the participants used a cameraphone with 4
of them using some other device as well (a DSLR or a point and shoot
camera).  All the participants took over 10 photos a week and used
photos in messaging services in addition to some photo sharing
websites.  The participants had a diverse employment background:
computer engineers, office managers, bartenders, and teachers.  There
was an even comic book and manga reading habit across the participants
(from infrequently to frequently).

\begin{figure}
  \centering
  \includegraphics[width=0.48\columnwidth]{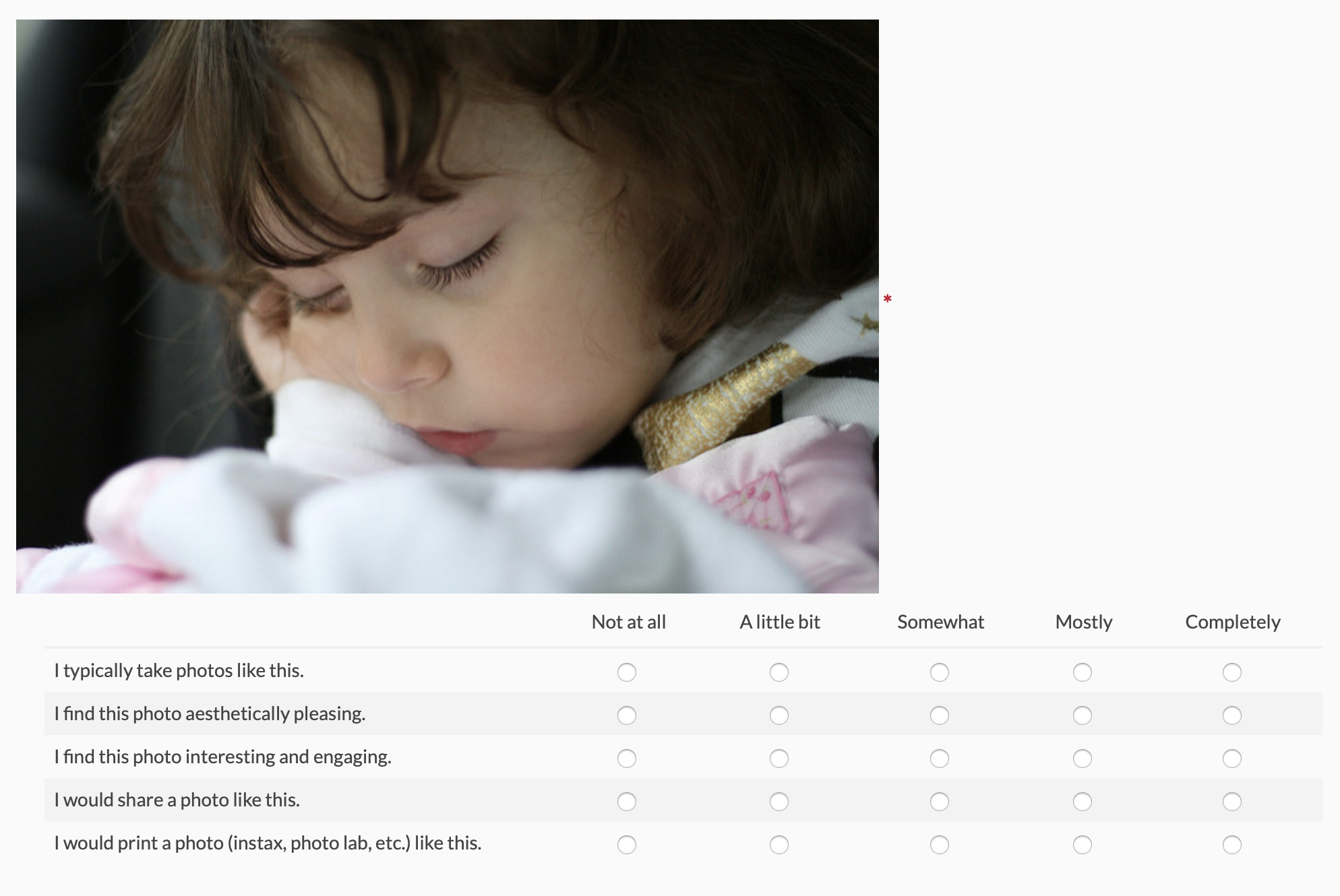}
  \caption{\label{fig:survey1} A question from the opening of a
    survey: a photo of a girl sleeping and a set of Likert questions
    asking if a user takes a photo like that or finds it aesthetically
    pleasing or engaging and if they would share this photo. The
    questions below are on a 5-point scale: if they typically take
    photos like this, if they find the photo aesthetically pleasing,
    if they find the photo interesting, and if they find they would
    share a photo like this.}
\end{figure}

The first section of the survey gathered information on the
participants.  Next each participant was shown a set of random photos
with a set of Likert 5-point scale questions asking if they typically
take photos like this, if they find the photo aesthetically pleasing,
if they find the photo interesting, if they would share a photo like
this, and if they would print a photo like this (via an instant
printer or photo lab). (see Figure~\ref{fig:survey1}) The instructions
for all three photo-question parts of the survey were to imagine they
took the photo themselves. Answers to ``I typically take photos like
this.'' were normally distributed least likely to most likely (6, 9,
13, 11, 6).  Aesthetic judgments were slightly different (11, 6, 10,
12, 6) and had more negatives which is likely due to selection of
everyday cameraphone images for the experiment.

\begin{figure}
  \centering
  \begin{subfigure}[t]{0.3\columnwidth}
    \includegraphics[width=\columnwidth]{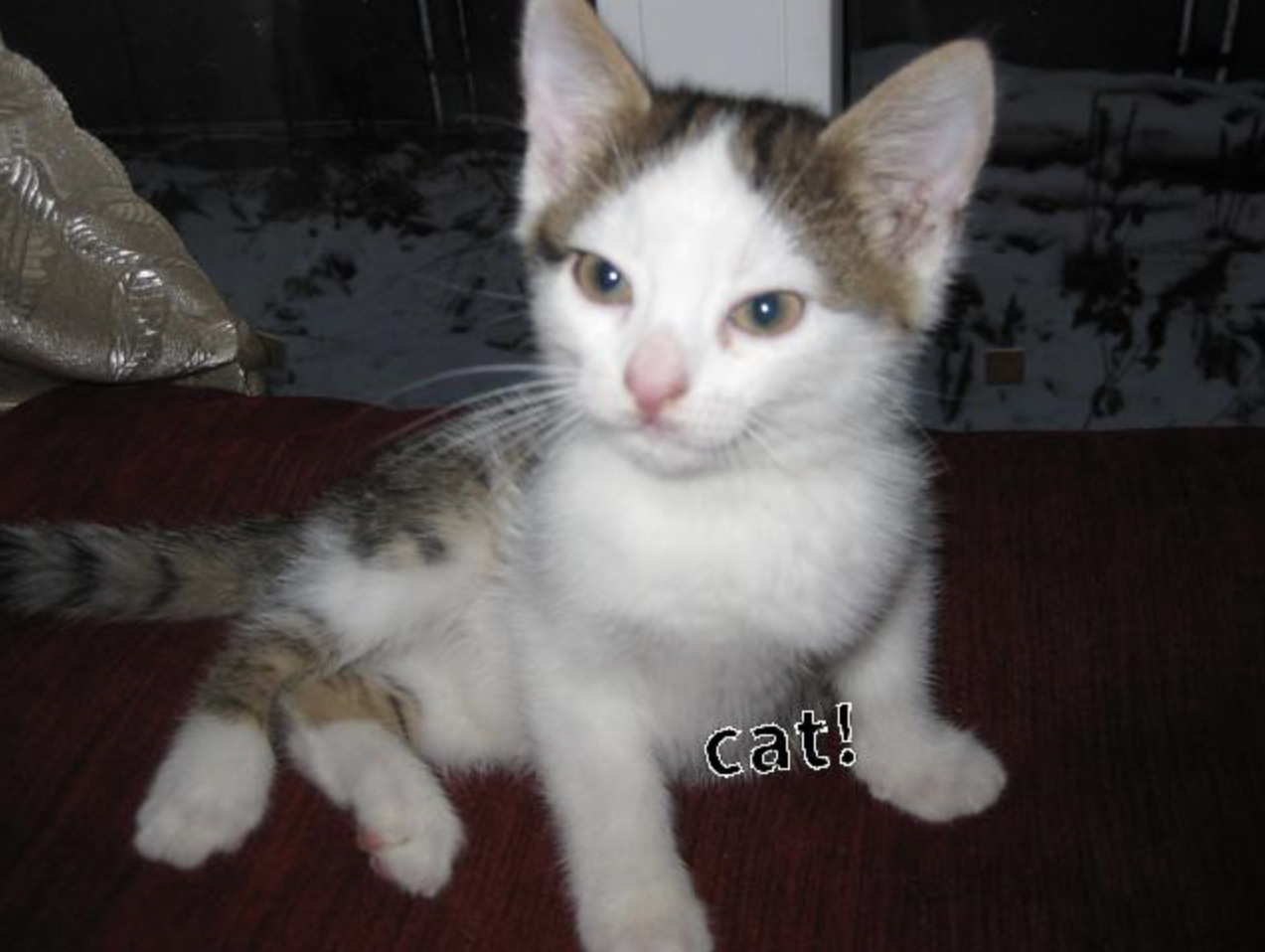}
  \caption{\label{fig:survey2a} An object based annotation.}
  \end{subfigure}
  \begin{subfigure}[t]{0.3\columnwidth}
    \includegraphics[width=\columnwidth]{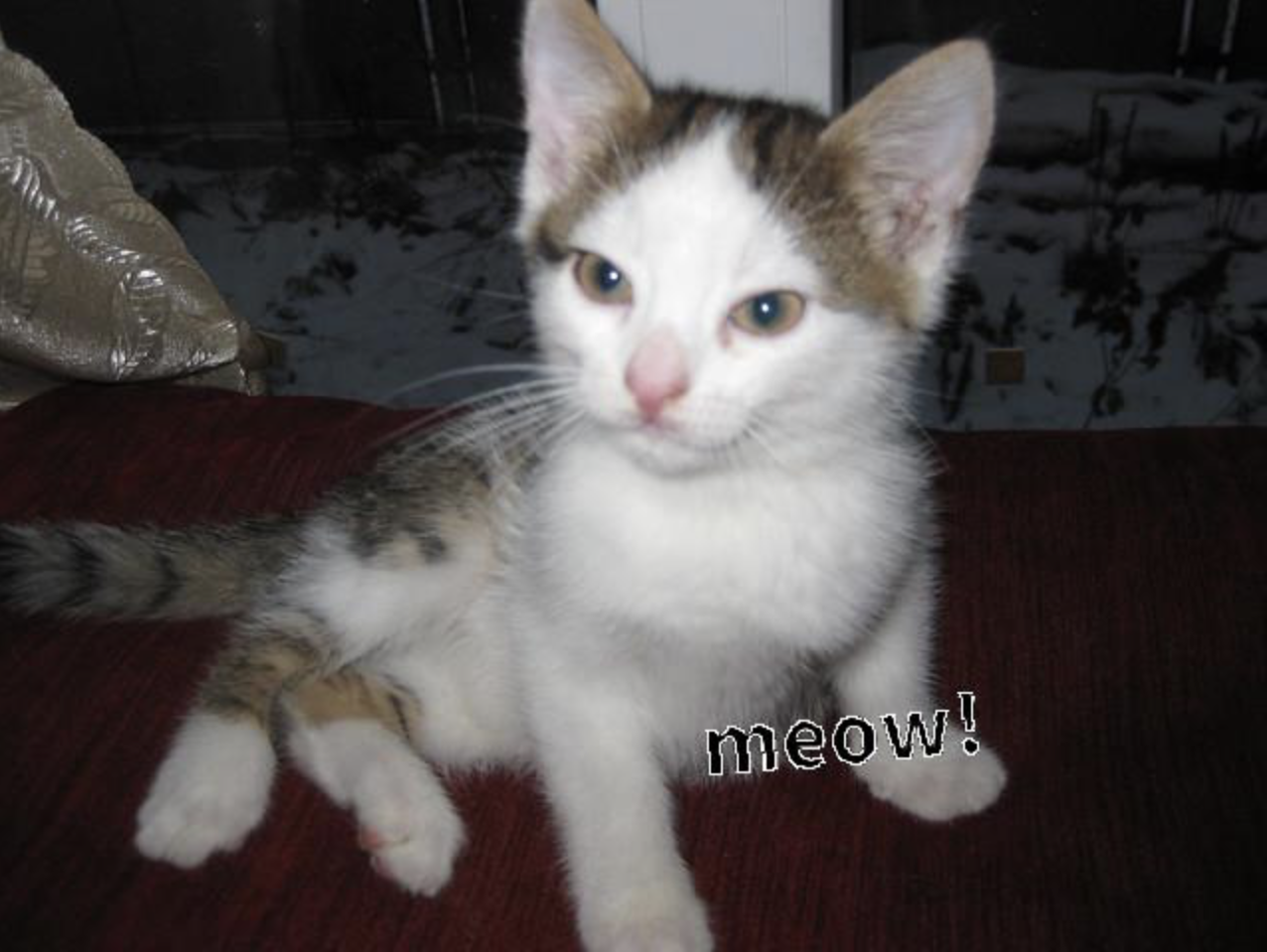}
    \caption{\label{fig:survey2b} A mimetic based annotation.}
  \end{subfigure}
  \caption{\label{fig:survey2} Following the baseline questions (see
    Figure~\ref{fig:survey1}), a subset of questions was asked about
    the same photos annotated by our Ideophone method and a standard
    object based label.}
\end{figure}

\begin{figure}
  \centering \includegraphics[width=.48\columnwidth]{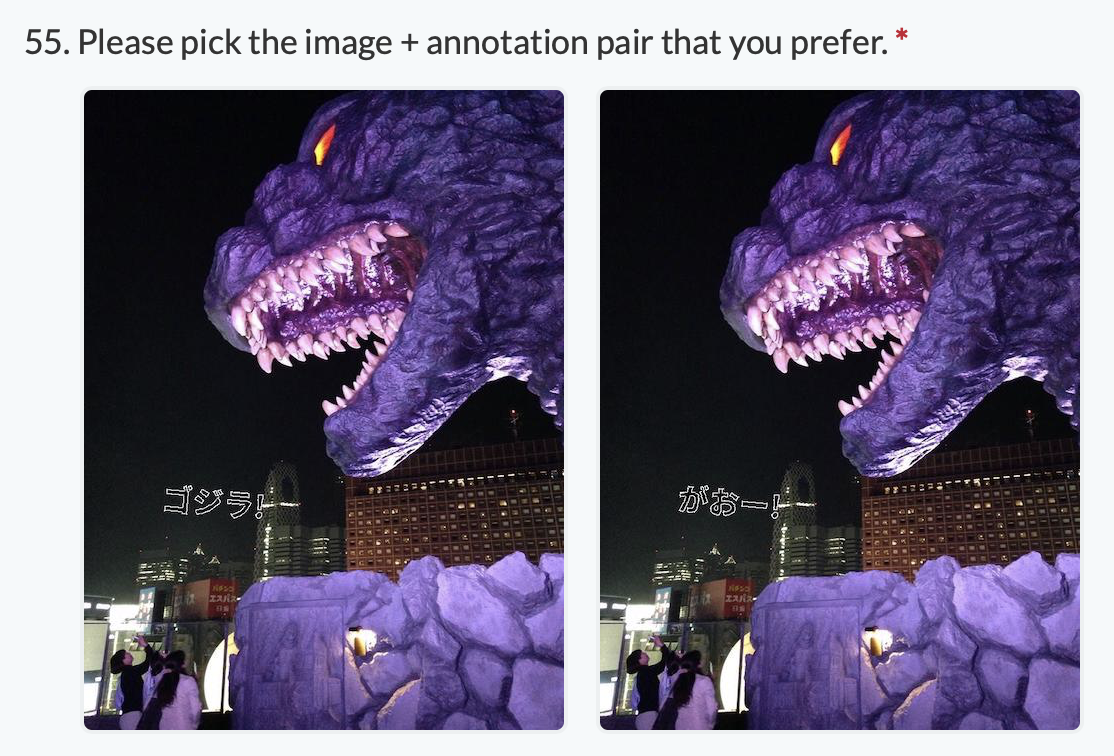}
    \begin{CJK}{UTF8}{min}
      \caption{\label{fig:survey4} The comparison section of the
        survey asked participants to pick between an object based and
        an ideophone based recommendation; these were presented
        randomly. Left: ゴジラ (Godzilla), Right:  がおー (Roar)}
  \end{CJK}
\end{figure}

We also asked of the un-annotated photos from the first section how
likely they were to engage with the photo via clicking like, sharing,
or printing the photo. For the next section we use the same photos
randomly presented with an object-based annotation and a mimetic-based
annotation from our method (Figure~\ref{fig:survey2}). For the purpose
of this exploration, we wish to see, by participant, is there an
increase or decrease in favor across those three dimensions.
Comparing the object-based annotations to the unannotated we see
10.5\% of the questions increased in score and 12.6\% decreased. The
ideophone-annotations increased by 17.0\% and decreased by 9.6\%. This
anecdotally shows an engagement increase for the ideophone
annotations, but more participants, and data, would be needed for a
further comparison and tests. The last comparison section asked
participants to pick an object versus ideophone annotation using new,
unseen photos (see Figure~\ref{fig:survey4}). These question and
choices were also randomized. The ideophone term annotations were
picked over the object terms at a ratio of $2.2 : 1$.

\begin{CJK}{UTF8}{min}
  Finally, there was a write in section asking what they liked and
  didn't like about the annotations. Most participants said the
  annotations (object and mimetic) were accurate; P1 expressed the
  accuracy of the matching given this survey comes from overseas.  The
  mimetic words were generally thought to ``reinforce the ideas in the
  viewers that the image wants to bring across.'' (P9) or, simply put,
  found joy in them ``Thank you for making me laugh several times.''
  (P8) The exception of a bowl of noodles ``Onomatopoeic is better
  than a noun, except `noodle' picture'' (P6); here the onomatopea was
  ずるずる or ``slippery'' in Japanese (which was in the dictionary as
  (1) being dragged and (2) ``eating soup noisily and without care for
  the surroundings'').  P7 stated the more aesthetic photos (in this
  case a shot of lightning in the night sky) should not be annotated
  at all as it detracts from the beauty of the image; this is
  congruent with past work on how people feel about filtering photos
  by Bakhshi et al.~\cite{bakhshi2015we}. Both negative comments about
  the annotations were seen in the study questions as well.  A few
  participants, P5, P6, and P10, thought the text should be rendered
  at a higher quality and composited with some transparency; this was
  an effect of building the system optimized for a 640$\times$480
  instant printer but did not surface during our survey testing
  period.  One participant (P9) stated their own images would affect
  their answers (they were requesting if that would be possible).
\end{CJK}


\section{Conclusion}
Ideophones are an overlooked area of automatic photo annotation. These
mimetic and onomatopoeic terms can be used to increase the engagement
on photos; we have shown how a semantic embedding can map visually
recognized objects to known ideophones via their definitions.  We find
improvements in visually compositing text is important but placement
was adequate in this limited study. A photo aesthetics classifier
could also be added as not to proactively modify high quality images.
Our system can also be extended to matching any set if items (terms,
graphics, etc.)  with a matching definition.  However, there exist
more opportunities for improvements in the matching system. Scene
graphs or automated captioning can be used to gather a better
understanding of the visual content for mapping.  Also, as most
cellphone cameras can capture video and stills simultaneously,
detecting which object is in motion from the video~\cite{Umeda_2012}
could further improve matching terms to salient objects.

\section{Acknowledgements}
Thanks to Takahiro Nori for translating the survey and to Nami
Tokunaga and Xiaojing Zhang for the early
testing. Figure~\ref{fig:survey2}'s cat photo \ccby\ \textsf{indamage}
on Flickr; all other images are author owned.

\small
\bibliographystyle{acm}
\bibliography{main.bib}

\end{document}